\newcommand{\beq}{\begin{eqnarray}}
\newcommand{\eeq}{\end{eqnarray}}

\documentclass[preprintnumbers,twocolumn,amsmath,amssymb,prl]{revtex4}

\usepackage{graphicx}

\begin{document}

\title{Spontaneous thermal runaway as an ultimate failure mechanism of materials}

\author{S. Braeck}
\email{simen.brack@fys.uio.no}
\author{Y.Y. Podladchikov}
\affiliation{Physics of Geological Processes (PGP), University of Oslo, P.O. Box 1048 Blindern,
N-0316 Oslo, Norway}

\begin{abstract}
The first theoretical estimate of the shear strength of a perfect
crystal was given by Frenkel~[Z. Phys. \textbf{37}, 572 (1926)]. He assumed that as slip occurred, two rigid
atomic rows in the crystal would move over each other along a slip plane. Based on this
simple model, Frenkel derived the ultimate shear strength to be about one tenth of
the shear modulus. Here we present a theoretical study showing that catastrophic
material failure may occur below Frenkel's ultimate limit as a
result of thermal runaway. We demonstrate that the condition for thermal runaway
to occur is controlled by only two dimensionless variables and, based on the thermal runaway
failure mechanism, we calculate the maximum shear strength $\sigma_c$ of viscoelastic materials.
Moreover, during the thermal runaway process, the magnitude of strain and temperature progressively
localize in space producing a narrow region of highly deformed material, i.e. a shear
band. We then demonstrate the relevance of this new concept for material failure known to occur
at scales ranging from nanometers to kilometers.
\end{abstract}

\maketitle

It is well known that the shear strength of real crystals
is typically several orders of magnitude smaller than Frenkel's ultimate shear strength limit.
This discrepancy is explained by the fact that real crystals contain defects such as
dislocations which lower the shear strength dramatically. Nevertheless, some materials,
like rocks in the Earth's interior and metallic glasses, apparently have strengths approaching
Frenkel's theoretical limit. For these materials it is reasoned that the mobility of the defects
is in one way or another reduced. For instance, the closure of cracks at high confining
pressures, non-planar crystal structure of minerals and disorder of mineral grain orientations
are all factors contributing to the high strength of rocks in the Earth's interior.
The high strength of metallic glasses is attributed to the high degree of structural disorder
causing dislocations to experience a large number of obstacles, reducing their
mobility and inhibiting plastic flow.

However, even if dislocation mobility is greatly reduced, materials subjected to
increasing loads do not necessarily fail according to Frenkel's model. When subjected to a high shear
stress which approaches, but is reproducibly lower than Frenkel's limit, these materials may fail by
deformation localized on a single or a few regions (shear bands) having thicknesses
that are orders of magnitude larger than interatomic spacing, but which are still very narrow
compared to the deforming sample size. Moreover, this extreme localization of shear is often
accompanied by extensive melting and resolidification of the material. This mode of
failure is manifested by the so-called pseudotachylytes in geological outcrops (cm thick deformation
bands believed to be the traces of large paleoearthquakes)~\cite{Andersen,Obata} and catastrophic failure
along ca. 10 nm thick shear bands reported in metallic glasses~\cite{Hays,Wright,Lewandowski}.

The occurrence of localized shear zones in the deeper parts of the earth's lithosphere
poses a problem due to its paradoxical nature. Earthquakes are usually attributed to the
intrinsic stick-slip character of brittle failure or frictional sliding on a pre-existing fault.
Brittle failure is expected as the primary mode of rock failure down to 15--20 kms, the so-called
seismogenic zone. Due to the high confining pressure at depth, however, these mechanisms alone are
not plausible explanations for earthquakes occurring significantly deeper than the seismogenic zone.
Several potential rock-weakening mechanisms have instead been proposed to facilitate
deep earthquakes, including structural changes such as dehydration embrittlement and olivine spinel
phase transition~\cite{Green} or thermal softening leading to shear instability~\cite{Griggs,Ogawa}.
Recent assessment of these models using indirect seismological evidence argues in favor of a
temperature-activated phenomenon, such as thermal shear instabilities, due to apparent temperature
dependence of deep earthquakes~\cite{Wiens}.
Similarly, structural strain softening and thermal softening are possible explanations for shear band
formation in metallic glasses~\cite{Molinari,Pampillo}.
A recent experiment~\cite{Lewandowski}\textbf, however, revealed that
the heated zone associated with an individual band was much wider
than the shear band itself. It was concluded~\cite{Lewandowski} that shear-band operation could not be fully
adiabatic and temperature rise could not therefore be the factor controlling shear-band thickness, thus
rejecting the thermal softening mechanism. In contrast, our work presented below shows that the non-adiabaticity of
the thermal softening process may in fact be the cause of strain-localization inside the hot zone, and we thus provide
an alternative interpretation to the experimental results in ref.~\cite{Lewandowski}.

It is well accepted that even below the conventional elastic limit, most real materials
show non-elastic rheological responses such as creep under constant load and stress relaxation under
constant extension induced by thermally excited defects and imperfections. In
accordance with these properties, most materials may be characterized as viscoelastic, i.e. the
rheology contains both viscous and elastic components.
Since the phenomena of creep and relaxation are thermally
activated processes, the viscosity is strongly temperature dependent (e.g. Arrhenius) and it is, in general,
a non-linear function of the shear stress~\cite{TS}.
The strong temperature dependence of the viscosity has important implications as it leads to thermal softening
of the material. Indeed, as first noted by Griggs and Baker~\cite{Griggs}, such a physical system is
inherently unstable: an increase in strain rate in a weaker zone causes a local temperature rise
due to viscous dissipation and weakens the zone even further. At high stresses, viscous dissipation
becomes substantial, and if heat is generated faster than it is conducted away, the local increase
in temperature and strain rate is strongly amplified. Under those conditions a positive feedback between
temperature rise and viscous dissipation is established and a thermal runaway develops.
To determine whether the thermal runaway mechanism can explain the aforementioned material failure, we approach
the problem by considering a simple viscoelastic model which accounts for the non-elastic behavior below
the ultimate yield point (Frenkel's limit). The temperature $T$ in the model is determined by the equation
for energy conservation which is coupled, through temperature dependent viscosity, to the rheology equation.

\begin{figure}[t!]
\begin{center}
\includegraphics[width=8cm]{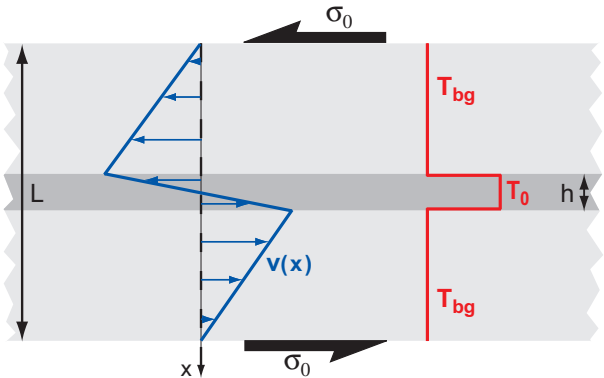}
\caption{Initial setup of the viscoelastic slab model discussed in the text
(cross-section in the $xy$-plane). The slab is in a state of stress of simple shear with zero velocity
($v=0$) boundary conditions. The shear stress $\sigma$, constant throughout the slab, initially ($t=0$)
equals the maximum value $\sigma_0$ and subsequently decreases with time due to relaxation and viscous
deformation in the interior. The blue and red lines show the initial velocity and temperature profiles,
respectively. The shaded region illustrates a small perturbation in temperature $T_0$ of width $h$
at the slab center. Elsewhere, the background temperature is $T_{bg}$. The geometry of the strain rate
profile concurs with that of the temperature profile.}
\label{fig:modelsetup}
\end{center}
\end{figure}

Our one-dimensional model (see fig.~\ref{fig:modelsetup}) consists of a viscoelastic slab of width $L$ at
initial temperature $T_{bg}$ except in the small central region having width $h$ and slightly elevated
temperature $T_0$. The boundaries are maintained at the temperature $T_{bg}$.
Our objective is to search for spontaneous modes of internal failure not aided or triggered
by the effect of additional far-field deformation. Hence, for time $t\geq 0$ we impose zero velocity $v$ at the
boundaries and assume that, without addressing the loading history ($t<0$), the slab initially ($t=0$) is
subjected to a shear stress $\sigma_0$. The shear stress $\sigma$ in the slab satisfies the equation for
conservation of momentum
\beq
\frac{\partial \sigma}{\partial x} = 0\,,
\label{eq:scndgoveq}
\eeq
which shows that $\sigma$ is independent of $x$ and hence only a function of the
time $t$. The viscoelastic rheology is represented by the Maxwell model~\cite{Malvern}, and is given by the
equation
\beq
\frac{\partial v}{\partial x} = \frac{1}{\mu(T,\sigma)}\sigma + \frac{1}{G}\frac{\partial\sigma}{\partial t}\,,
\label{eq:Maxwellrheology}
\eeq
where $v(x,t)$ is the velocity, $G$ is the constant shear modulus and $\mu(T,\sigma)$ is the viscosity.
The dependence of $\mu$ on $T$ and $\sigma$ may be written as
\beq
\mu(T,\sigma)=A^{-1}e^{E/RT}\sigma^{1-n}\, ,
\label{eq:viscosity}
\eeq
where $A$ and $n$ are constants, $E$ is the activation energy and $R=8.3$ JK$^{-1}$mole$^{-1}$ is the
universal gas constant.
Since $\sigma(t)$ is independent of $x$, it follows from eq.~(\ref{eq:Maxwellrheology}) that the geometry
of the strain rate ($\partial v/\partial x$) profile at any instant concurs with that of the temperature profile
$T(x,t)$. Utilizing the zero velocity boundary condition, equation~(\ref{eq:Maxwellrheology}) may
be integrated to obtain the equation which governs the time-dependence of $\sigma$:
\beq
\frac{\partial \sigma}{\partial t} &=& -\frac{GA}{L}\sigma^n
\int_{-\frac{L}{2}}^{\frac{L}{2}}e^{-\frac{E}{RT}}dx\, .
\label{eq:finstresseq}
\eeq
The temperature is determined by the equation for energy conservation
\beq
\frac{\partial T}{\partial t} &=& \kappa \frac{\partial^2 T}{\partial x^2}
+ \frac{A}{C}\sigma^{n+1}e^{-\frac{E}{RT}}\, ,
\label{eq:fintempeq}
\eeq
where the last term accounts for viscous dissipation in the system.
Equations~(\ref{eq:finstresseq}) and (\ref{eq:fintempeq}) constitute a closed system
of coupled ordinary and partial differential equations for two unknown functions
$\sigma(t)$ and $T(x,t)$.
\begin{figure*}[t!]
\begin{center}
\includegraphics[width=8cm]{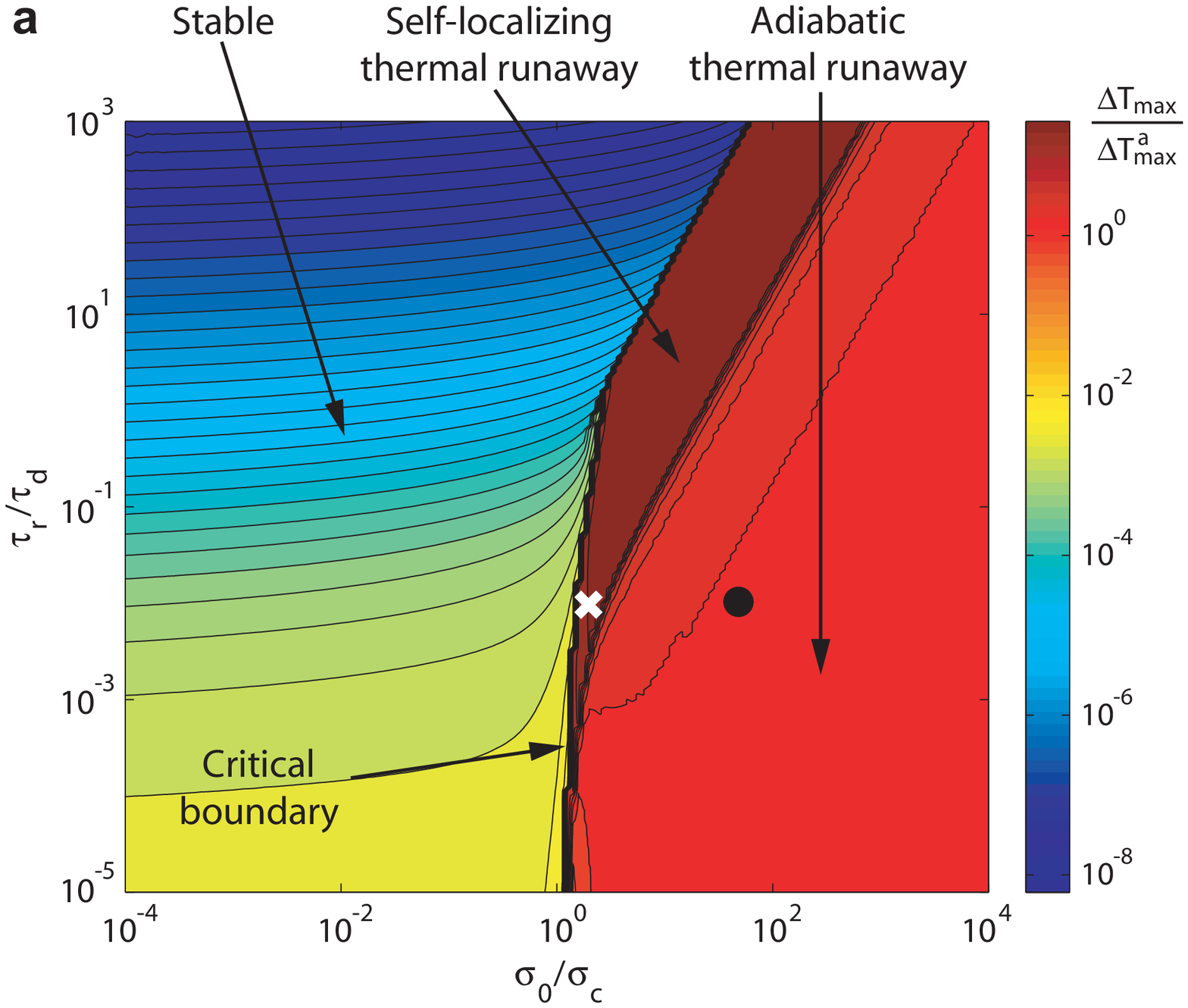} \hspace{1cm}
\includegraphics[width=8.5cm]{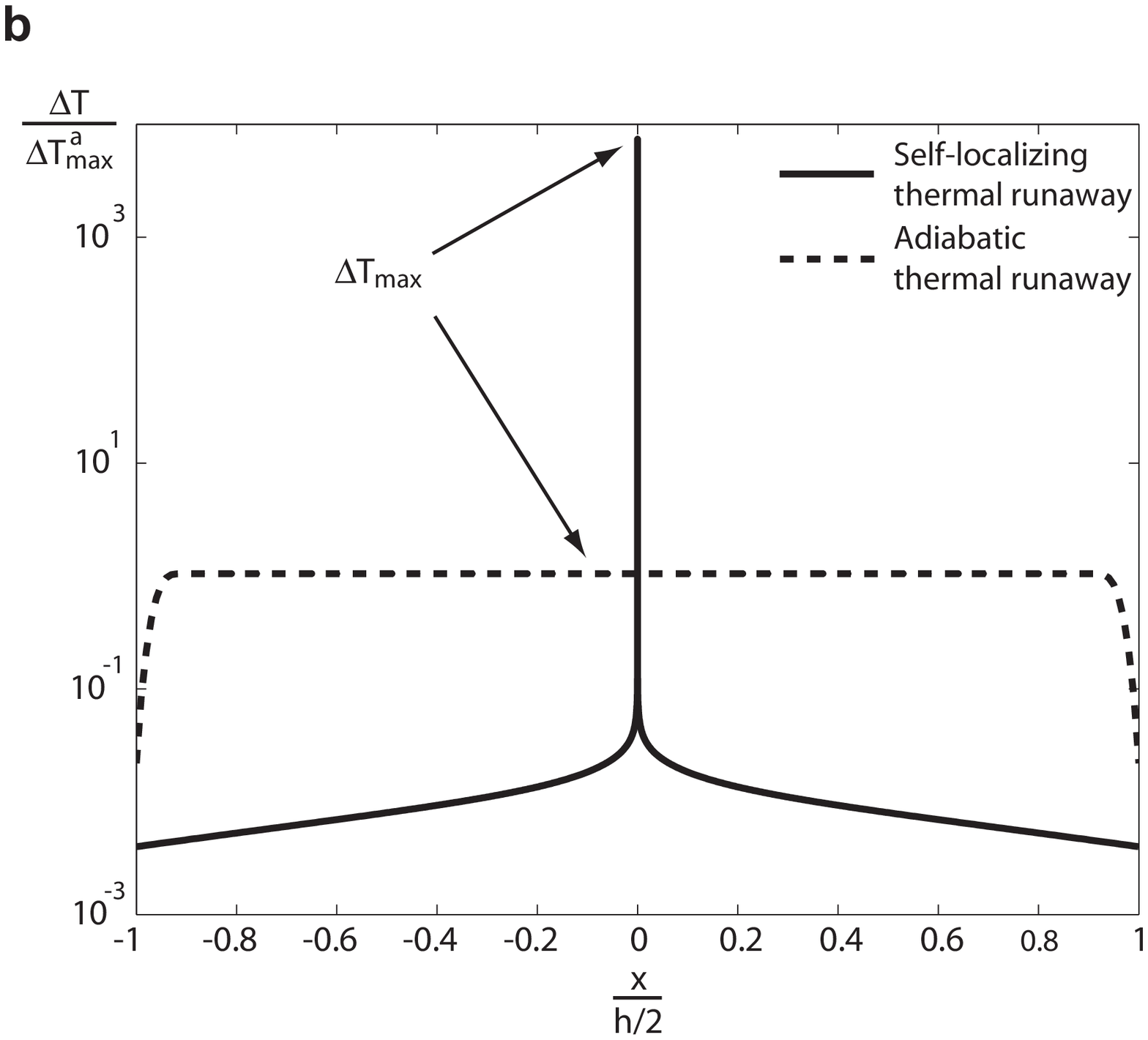}
\caption{Dependence of the maximum temperature rise $\Delta T_{max}$ on the dimensionless
variables $\sigma_0/\sigma_c$ and $\tau_r/\tau_d$.
\textbf{a}, Contour plot of $\Delta T_{max}$ scaled by the adiabatic temperature rise
$\Delta T_{max}^a$ as a function of  $\sigma_0/\sigma_c$ and $\tau_r/\tau_d$. The dark lines are
contour lines. Due to the computational effort of fully resolving the entire self-localizing thermal runaway
processes, the temperature at the very late stages of these processes are not presented in this plot.
\textbf{b}, Profiles of the temperature rise $\Delta T = T-T_0$ inside the initially perturbed
zone $|x|\leq h/2$ at the time when the maximum temperature is reached. The solid line illustrates
the self-localizing thermal runaway process at the location of the cross in \textbf{a}. The dashed
line illustrates the adiabatic thermal runaway at the location of the dot in \textbf{a}.}
\label{fig:Phasediagram}
\end{center}
\end{figure*}

First, to determine the conditions necessary for thermal runaway to occur, a linear stability analysis was carried out.
Equation~(\ref{eq:finstresseq}) was approximated for the initial stages by substituting the initial conditions
for $T$ and $\sigma$. This yields a characteristic time $\tau_r=\mu_0/(2G\Delta_p)$ for stress relaxation.
Here $\mu_0\equiv \mu(T_0,\sigma_0)$ and $\Delta_p=h/L + e^{E/RT_0-E/RT_{bg}}$ is a factor which characterizes
the initial perturbation. Linearization of the temperature equation with account of stress relaxation yields that the
growth of the perturbation in the initial stages is controlled by the two dimensionless variables
$\tau_r/\tau_d$ and $\sigma_0/\sigma_c$, where the thermal diffusion time $\tau_d=h^2/\kappa$ ($\kappa$ is the thermal
diffusivity) and the newly introduced stress
\beq
\sigma_c = \sqrt{2\Delta_p\frac{GCR}{E}}\,T_0 \,.
\label{eq:criticalstress}
\eeq
Here $C$ denotes the heat capacity per volume. The solution to the linearized temperature equation
is found to be unstable if $\sigma_0/\sigma_c > f(\tau_r/\tau_d)$. In the limit $\tau_r/\tau_d << 1$, which
corresponds to near adiabatic conditions, the function $f$ quickly approaches the lower bound
$f\approx 1$ and the solution therefore becomes unstable if $\sigma_0 > \sigma_c$. Thus $\sigma_c$ is the
critical stress above which a thermal runaway may occur and therefore provides an estimate of the maximum
shear strength of viscoelastic materials. Initial stages of thermal runaway instability in a more general
setup was recently investigated in ref.~\cite{Boris}, including two-dimensional verification
of the one dimensional predictions. The results of our linear stability analysis are in agreement with these
numerical estimates in the limit of vanishing boundary velocity.

Non-linear evolutions of the unstable runaway modes rapidly deviate from the exponential growth in time
predicted by linear analysis. Since important information about the deformation process can be inferred from
the increase in temperature, we choose the maximum temperature rise $\Delta T_{max} = T_{max}-T_0$ during the
deformation process as our main physical quantity to study. This enables us to quantify even the later stages of
the thermal runaway process not considered in the linear analysis. A simple estimate of $\Delta T_{max}$ during
thermal runaway may be obtained assuming adiabatic conditions. In this case all the elastic energy in the system
is uniformly dissipated as heat in the perturbed zone and overall energy balance yields the adiabatic
temperature rise
\beq
\Delta T_{max}^a = \frac{L\sigma_0^2}{2hGC} \,.
\label{eq:adtemprise}
\eeq
Guided by these analytical estimates, the complete time evolution of $T$ and $\sigma$ was subsequently
investigated by numerical methods. We simplified the problem by dimensional analysis reducing it from
one containing thirteen dimensional parameters to one containing six dimensionless parameters.
The dimensionless form of the coupled set of equations (\ref{eq:finstresseq}) and (\ref{eq:fintempeq})
were solved numerically using a finite-difference method with non-uniform mesh and a tailored variable
time-step in order to resolve the highly non-linear effect of localization.
We have systematically varied all six dimensionless parameters and computed $\Delta T_{max}$ for each
temperature evolution. Remarkably, it is possible to present $\Delta T_{max}$ normalized by the
adiabatic temperature rise~(eq.~(\ref{eq:adtemprise})) as a function of only two combinations of parameters,
namely $\sigma_0/\sigma_c$ and $\tau_r/\tau_d$, as previously suggested by the linear stability analysis.
A representative set of runs is shown in fig.~\ref{fig:Phasediagram}a.
This ``phase diagram'' was computed by varying two of the dimensionless parameters and fixing the remaining
four. The plot exhibits a low-temperature region corresponding to stable deformation processes, and a
high-temperature region corresponding to thermal runaway processes. These regions are sharply
distinguished by a critical boundary having a location that correlates well with stability-predictions
of the linear analysis. The phase diagram is ``representative'' in that it is insensitive to which
two out of the six dimensionless parameters are varied, keeping the remaining four fixed.

In the neighborhood of the critical boundary in the high-temperature region,
however, the temperatures are found to be much larger than the adiabatic temperature rise
$\Delta T_{max}^a$. In this region we observe a continuous localization of the temperature and strain
profiles during the deformation process, i.e. the runaway is spatially self-localizing. This localization
effect is illustrated for the temperature profile in fig. 2b.
The elastic energy is thus dissipated in a zone much narrower than
the width of the initial perturbation resulting in much larger temperatures. The self-localization
of the runaway process arises from the effects of thermal diffusion: by diffusion the temperature
profile acquires a peak in the center where the effect of the positive feedback mechanism accordingly is
maximized. The runaway therefore accelerates faster in the center than in the regions outside and
the deformation process finally terminates in a highly localized shear band with a characteristic width much
smaller than the characteristic width $h$ of the initial perturbation.

To evaluate the relevance of thermal runaway as a potential failure mechanism in nature, we now
consider two case examples comparing critical stress for thermal runaway with Frenkel's ultimate
shear strength limit ($\sigma\sim G/10$). First, we estimate the condition necessary for thermal runaway to
occur in olivine-dominated mantle rocks. For olivine (see ref. \cite{TS}) $G=7\times 10^{10}$ Pa,
$C=3\times 10^6$ Jm$^{-3}$K$^{-1}$ and $E=5,23\times 10^5$ Jmole$^{-1}$, while we assume $T_{bg}=700$ K
(corresponding to a depth of about $40$ km), $T_0$ = 701--720 K and $h=10$ m, $L=10$ km. Substituting these values in
equation~(\ref{eq:criticalstress}) we predict that thermal runaway possibly occurs if the stress exceeds a critical
value in the range $\sigma_c$ = 0.5--1.7 GPa, i.e. $\sigma_c$ = $G$/140--$G$/40. This estimate agrees rather well
with the values observed in experimental studies of rock deformation under high confining pressure, where typical
failure stresses are in the range 0.5--2 GPa~\cite{Griggs,Renshaw}.
Second, for a metallic glass, typical values are $C=2,6\times10^6$ Jm$^{-3}$K$^{-1}$ (ref. \cite{Lewandowski}),
$G=34$ GPa (ref. \cite{Johnson}) and $E$ = 100--400 kJmole$^{-1}$ (ref. \cite{Wang}
and \cite{Chen}). Assuming $h=1$ $\mu$m, $L=1$ cm, $T_{bg}=300$ K (room temperature) and $T_0=301$ K
we obtain a critical stress in the range $\sigma_c$ = 0.4--1.1 GPa, i.e. $\sigma_c$ = $G$/90--$G$/30. For comparison,
we note that the shear yield strength of the metallic glass Vitreloy 1 is about 0.8 GPa = $G$/40~\cite{Lewandowski,Johnson},
a value consistent with our estimate of the critical stress necessary for thermal runaway.

The magnitude of the critical stress in these two estimates is remarkably similar considering the
very different types of materials discussed and the enormous difference in time and length scales. However, as is
evident from eq.~(\ref{eq:criticalstress}), the quantities which govern the kinetic processes in these systems
(in particular the scale $h$ and the poorly constrained viscosity $\mu$)
do not appear in the expression for $\sigma_c$. It is not surprising, therefore, that the stress required to
initiate spontaneous thermal runaway is relatively well constrained to about 1 GPa.

These estimates and the concept of self-localizing thermal runaway demonstrate that our simple
model is sufficient to explain failure below Frenkel's ultimate shear strength limit and strain
localization at scales much larger than the lattice spacing. The fact that initiation of thermal runaway
depends so weakly on the kinetic quantities gives confidence in the application of
such models even to the very large scales involved in continental
deformation~\cite{Boris,Regenauer-Lieb,Regenauer-Lieb-Yuen,Kameyama-Yuen}.

The authors thank K. Mair (PGP) and S. Medvedev (PGP) for discussions.
This work was supported by the Norwegian Research Council through a Center of Excellence
grant to PGP.

\end{document}